\documentclass{PoS}

\newcommand{\be}{\begin{equation}}
\newcommand{\ee}{\end{equation}}
\newcommand{\bea}{\begin{eqnarray}}
\newcommand{\eea}{\end{eqnarray}}

\title{The Fate of the Higgs Vacuum}

\ShortTitle{The Fate of the Higgs Vacuum}

\author{\speaker{Ruth Gregory}%
\thanks{Supported by STFC (Consolidated Grant ST/J000426/1). Preprint no: DCPT-16/47.}
\thanks{Also supported by Perimeter Institute. Research at Perimeter 
Institute is supported by the Government of
Canada through Industry Canada and by the Province of Ontario through the
Ministry of Research and Innovation.}\\
Centre for Particle Theory, Durham University,
South Road, Durham, DH1 3LE, UK\\
Perimeter Institute, 31 Caroline Street North, Waterloo, 
ON, N2L 2Y5, Canada\\
        E-mail: \email{r.a.w.gregory@durham.ac.uk}}

\author{Ian G. Moss$^\dagger$\\
School of Mathematics and Statistics, Newcastle University,
Newcastle Upon Tyne, NE1 7RU, U.K.\\
        E-mail: \email{ian.moss@newcastle.ac.uk}}

\abstract{This talk reviews our recent work showing how tiny black holes can act as nucleation
sites for the decay of the metastable Higgs vacuum \cite{GMW,BGM1,BGM2,BGM3}. 
We start by discussing the formation of thin wall bubbles of true vacuum inside a false 
vacuum, and show how adding a black hole lowers the action of the Euclidean tunneling
solution, thus strongly enhancing the probability of vacuum decay. We then review
numerical results for the Higgs vacuum showing that the decay rate is even higher 
for these ``thick wall'' bubbles. The results imply either tiny black holes are not a 
component of our universe, or BSM corrections to the Higgs potential must stabilise 
our vacuum.}

\FullConference{38th International Conference on High Energy Physics\\
		3-10 August 2016\\
		Chicago, USA}

\begin{document}

\section{Introduction}

One of the more curious aspects about discovery of the Higgs 
\cite{ATLAS:2012ae,Chatrchyan:2012tx}, is that its mass suggests
that our universe is metastable. The running of the Higgs 
self-coupling indicates the true Higgs vacuum lies at large
expectation values of the Higgs and negative vacuum energy
\cite{Degrassi:2012ry}.
Although a metastable vacuum is somewhat disconcerting, the key
factor is the lifetime for its decay: a region of the universe must
tunnel through a sizeable energy barrier, with a typical
probability dominated by an exponential factor \cite{Kobzarev:1974cp} 
\begin{equation}
\Gamma \sim A e^{-B/\hbar}
\label{decayaction}
\end{equation}
where $B$ is the action of a solution to the Euclidean field
equations interpolating from the metastable (false) to the
true vacuum. If this action is large, then the 
lifetime of our vacuum can be many orders of magnitude
greater than the age of our universe, and therefore not
necessarily a problem.

The Euclidean solution, or ``bounce'', was analysed
by Coleman and collaborators \cite{Coleman:1977py,Callan:1977pt,Coleman:1980aw},
whereby the decay was understood as bubble nucleation.
The idea is that a bubble of true vacuum forms within the false, and
typically there is an energy balance between the `cost' of the bubble wall
(the energy barrier between false and true vacua) and the `gain' from the
interior of the bubble now being at lower energy. Optimising this
energy pay-off gives the critical bubble size that corresponds to the
Euclidean solution -- the instanton -- that drives vacuum decay. Once
formed, the bubble expands, and we have a first order
phase transition from the false to the true vacuum.

This picture, while intuitive, is incomplete, as we must take into account gravity:
a false vacuum energy will
gravitate, and we must now consider bubbles between false vacua of one cosmological
constant to true vacua of a lower cosmological constant. In \cite{Coleman:1980aw},
Coleman and de Luccia (CDL) did precisely this. It turns out that gravity completely
fixes the bubble radius from the energy of the wall. The CDL instanton action is simply
a curved space generalisation of the flat space bubble.

All of these calculations however are rather idealised, they refer to a single bubble in a
completely pure and featureless universe. In reality however, our universe is not 
featureless, and phase transitions are rarely clean, indeed, they are often catalysed 
by the presence of an impurity. How dependent are the results of Coleman et al.\
on the assumptions of homogeneity and isotropy? Here, we briefly review our
work \cite{GMW,BGM1,BGM2,BGM3}
exploring this issue by introducing a simple gravitational impurity,
a black hole, and showing how even a single tiny black hole can overturn our
picture of how stable our universe is.

\section{Thin Wall Tunneling}

In thin wall tunneling \cite{Coleman:1980aw,GMW,BGM1,BGM2},
the physical input is that we have a potential with two local mimima 
(false and true vacuum) with a sufficient energy barrier that a Euclidean solution 
interpolating between the two will be very thin in comparison to the radius of the 
instanton bubble so that we can use the Israel formalism \cite{Israel}. While this
requires a putative quantum gravity correction to the Higgs potential to ensure a
second, stable, true vacuum (see next section) it allows us to easily
analytically explore the impact of a black hole, and provides a proof of
principle of black hole catalysis of decay.

In \cite{Coleman:1980aw}, CDL showed how to construct the
gravitational version of Coleman's thin wall calculation \cite{Coleman:1977py}
that extracted the key physics from the decay process. In gravity, provided
we have enough symmetry, we can solve the Euclidean Einstein equations
for a thin wall analytically, and CDL found an explicit expression for the action 
of a bubble between different vacua in terms of $\sigma$, the energy per unit 
area of the wall, and ${\cal E}$, the difference in vacuum energy.

Adding a black hole to the system turns out to be straightforward, the relevant
results for black holes and walls were derived in \cite{BCG}, which proved that
the general solution is a bubble wall separating two black hole spacetimes
\be
ds^2 = f_\pm(r) d\tau_\pm^2 + \frac{dr^2}{f_\pm(r)} + r^2 d\theta^2 + r^2 \sin^2\theta d\varphi^2
\quad {\rm where}\quad
f_\pm(r) = 1 - \frac{2GM_\pm}{r} + \frac{\Lambda_\pm r^2}{3}
\ee
Inside the bubble, we have a vacuum energy ${\cal E}_- = \Lambda_-/8\pi G$,
and possible a remnant black hole of mass $M_-$. Outside the bubble, the
vacuum has energy $\Lambda_+/8\pi G$, and $M_+$ is the mass of
the seed black hole catalysing the decay. The bubble lies at some $r=R(\tau)$,
satisfying a Friedmann-like equation
\be
\left ( \frac{\dot R}{R} \right)^2 = (2\pi G\sigma)^2 - \frac{f_++f_-}{2R^2}
+ \frac{(f_+-f_-)^2}{16 R^4 (2\pi G\sigma)^2}\;.
\ee
For very small mass black holes, the solution is time dependent,
and the lowest action instanton has no remnant black hole -- the instanton is thus a 
perturbed CDL bubble. For larger seed masses however the nature of the instanton 
changes, and the bubble is now `static' and contains a remnant black hole.

To find the action, we have to calculate the difference between the background black hole 
solution and the black hole with bubble solution. One of the key technicalities resolved 
in \cite{GMW} is the treatment of conical singularities that can arise in the Euclidean solution, we
refer the reader to \cite{GMW} for full detail. 
The static tunnelling instanton is the one  which is most
relevant for vacuum decay of our universe, and for this case a remarkable 
simplification results in the action depending only on the areas of the event horizons,
\begin{equation}
B=\frac{{\cal A}_+}{4G}-\frac{{\cal A}_-}{4G}
\end{equation}
We recognise these terms as the black hole entropies, and the vacuum decay 
rate is therefore consistent with the entropy-fluctuation formula first 
discovered by Einstein: $\Gamma\propto e^{\Delta S}$.
The rate is considerably larger than the CDL vacuum tunnelling rate, 
however, there is another quantum decay 
channel for a small black hole: Hawking evaporation.

Black holes emit Hawking radiation, and in consequence lose mass, leading
to a finite lifetime of order $\Gamma_H\approx 3.6\times 10^{-4}(G^2 M_+^3)^{-1}$
\cite{Page:1976df}. 
To estimate our vacuum decay, we need both the instanton action and an
estimate for the prefactor $A$.
It turns out that for seed masses sufficiently above the Planck scale that
we trust our approximations, we are well into the static instanton branch,
for which the action simplifies to the entropic form: $B =
\pi (r_+^2 - r_-^2)/G$, and following Callan and Coleman, we determine $A$
by taking a factor $(B/2\pi)^{1/2}$ for the translational zero mode of the
instanton, estimating the determinant piece at $(GM_+)^{-1}$ 
by dimensional analysis. Putting together, we obtain the branching ratio
for tunneling over evaporation as:
\be
\frac{\Gamma_D}{\Gamma_H} \approx 40 \frac{M_+^2}{M_p^2} 
\sqrt{B} e^{-B}
\label{branchratio}
\ee
This equation will be a key factor in the determination of the relevance of
vacuum tunneling. It is correct whether or not the thin wall approximation is
used, and simply requires knowledge of the bounce action, $B$. For the thin
wall, it turns out (roughly) that $B \propto M_+/M_p$, leading to a branching ratio
greater than one for a range of thin wall data. Whether or not this is relevant to
the Higgs potential we now determine.

\section{Higgs metastability}

In order to decide whether enhanced vacuum decay is relevant for
the Higgs, we must explore the instantons for the actual Higgs potential. 
This requires a full numerical analysis of the instantons for a range of 
parameter space relevant to the standard model (SM), and we now review
\cite{BGM1,BGM3}, first discussing our modelling of the running of the coupling,
then the bubble solutions. The high energy effective potential for the Higgs field 
within the standard model has been determined by a two-loop calculation 
\cite{Degrassi:2012ry},
and is conventionally written in terms of an effective coupling, as
$
V(\phi)=\frac14\lambda_{\rm eff}(\phi)\phi^4.
$
\begin{figure}[htb]
\begin{center}
\includegraphics[scale=0.395]{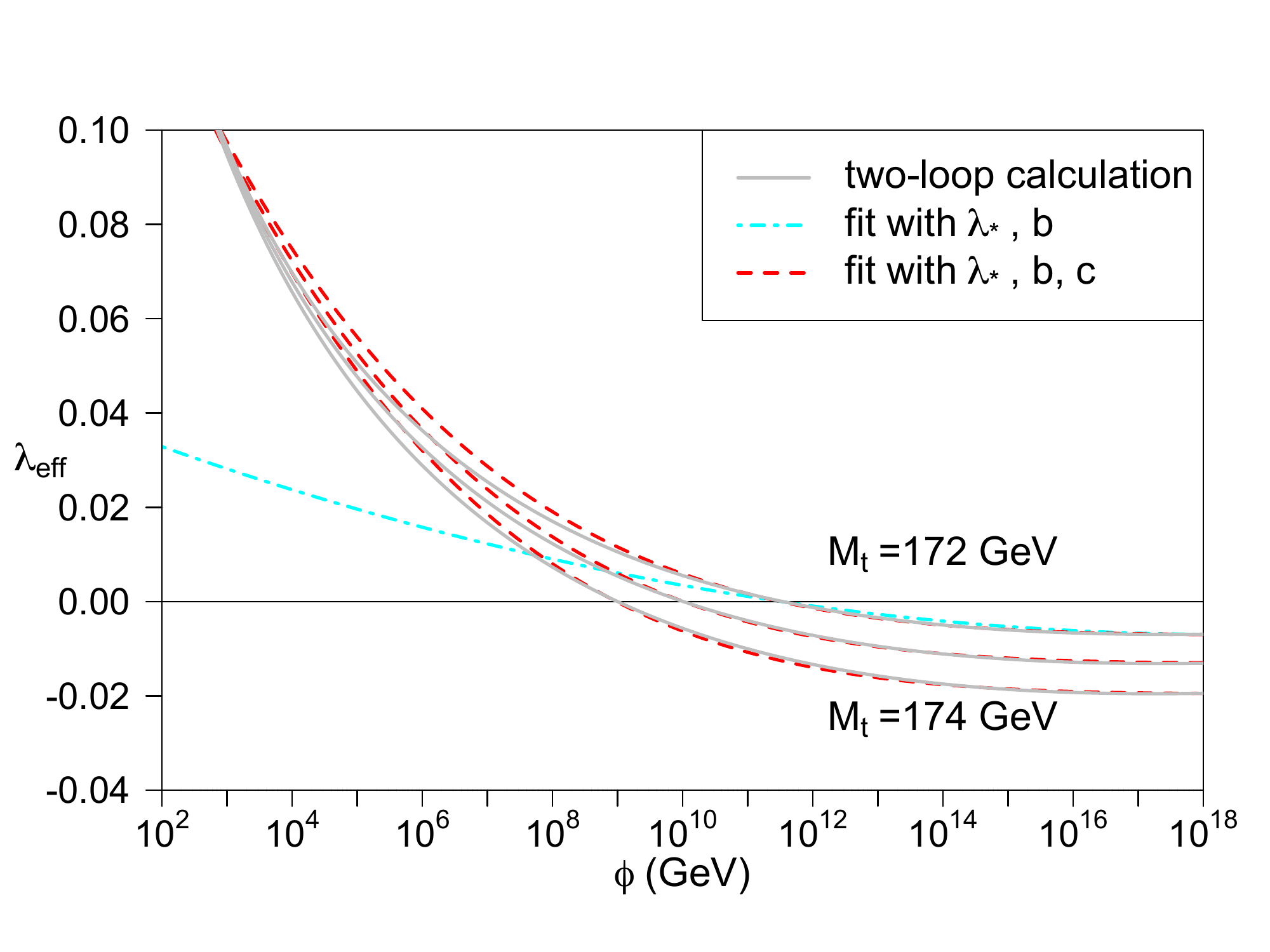}
\includegraphics[scale=0.35]{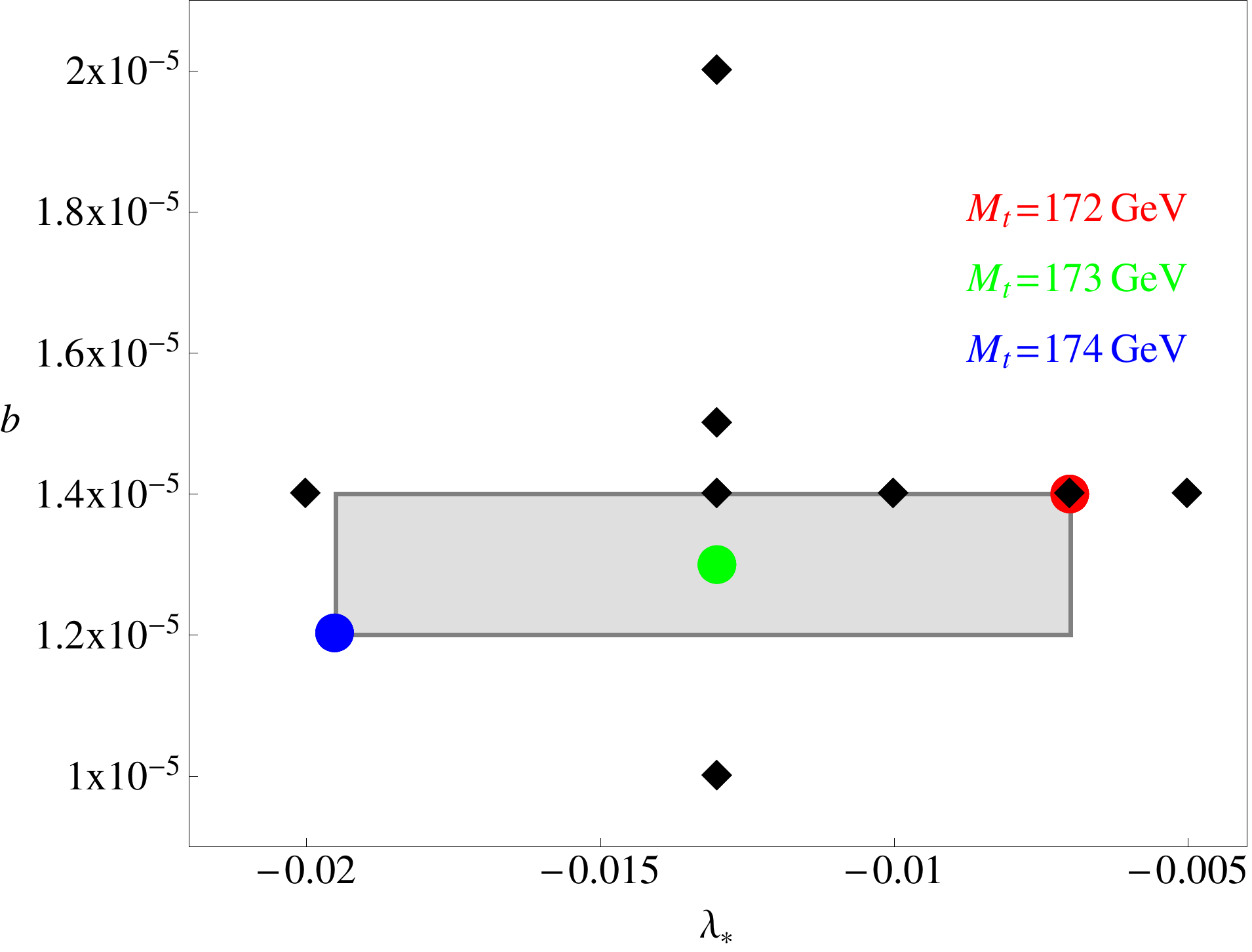}
\caption{
Left: The analytic modelling of the Higgs coupling demonstrating the fit with the
two-loop calculation over a wide range of scales. 
Right: The parameter space explored numerically. The shaded box representing
the SM range, and diamond markers the specific parameter values for we used.
}
\label{fig:lambda}
\end{center}
\end{figure}
The main uncertainty in the potential comes from the top quark mass uncertainty,
and for given $m_t, m_H$ can be computed by direct numerical integration of 
the $\beta-$functions \cite{Degrassi:2012ry}. Rather than compute the precise
potential for each possible value of $(m_t, m_H)$, we instead take an analytic 
three-parameter fit to the potential 
\begin{equation}
\lambda_{\rm eff}(\phi)=\lambda_*+b\left(\ln{\phi\over M_p}\right)^2
+c\left(\ln{\phi\over M_p}\right)^4.
\label{higgspot}
\end{equation}
that gives a much better fit over the large range of $\phi$ relevant for 
tunnelling phenomena, and allows us
to explore parameter space beyond the standard model. In practise, 
we fix the value of $\lambda_{\rm eff}$ at the electroweak scale,
which leaves two fitting parameters that we take to be $\lambda_*$ 
and $b$. Figure \ref{fig:lambda} shows the fit to $\lambda_{\rm eff}$,
and the range of parameter space we explore relative to the standard 
model. 

\begin{figure}
%
\includegraphics[scale=0.27]{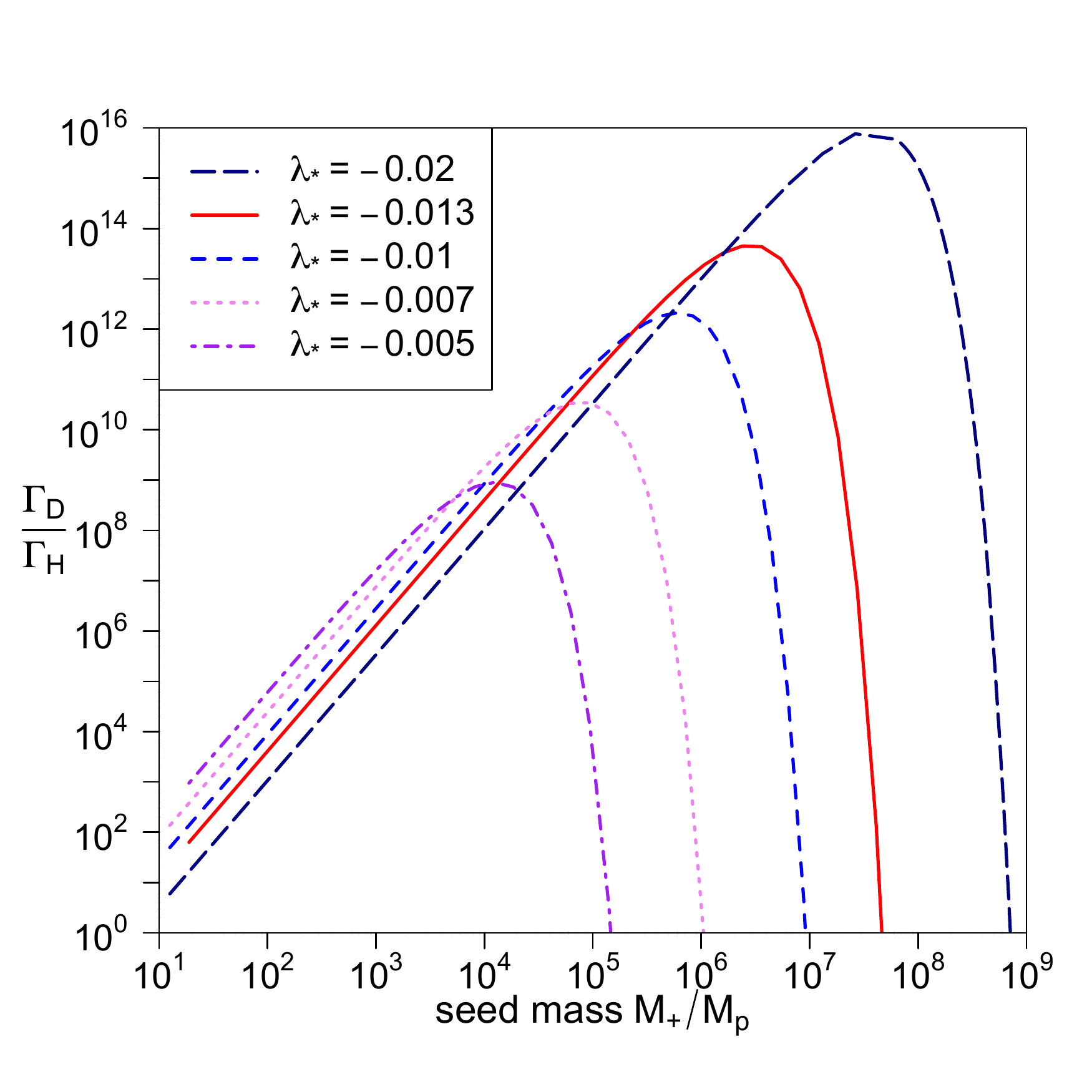}~
\includegraphics[scale=0.27]{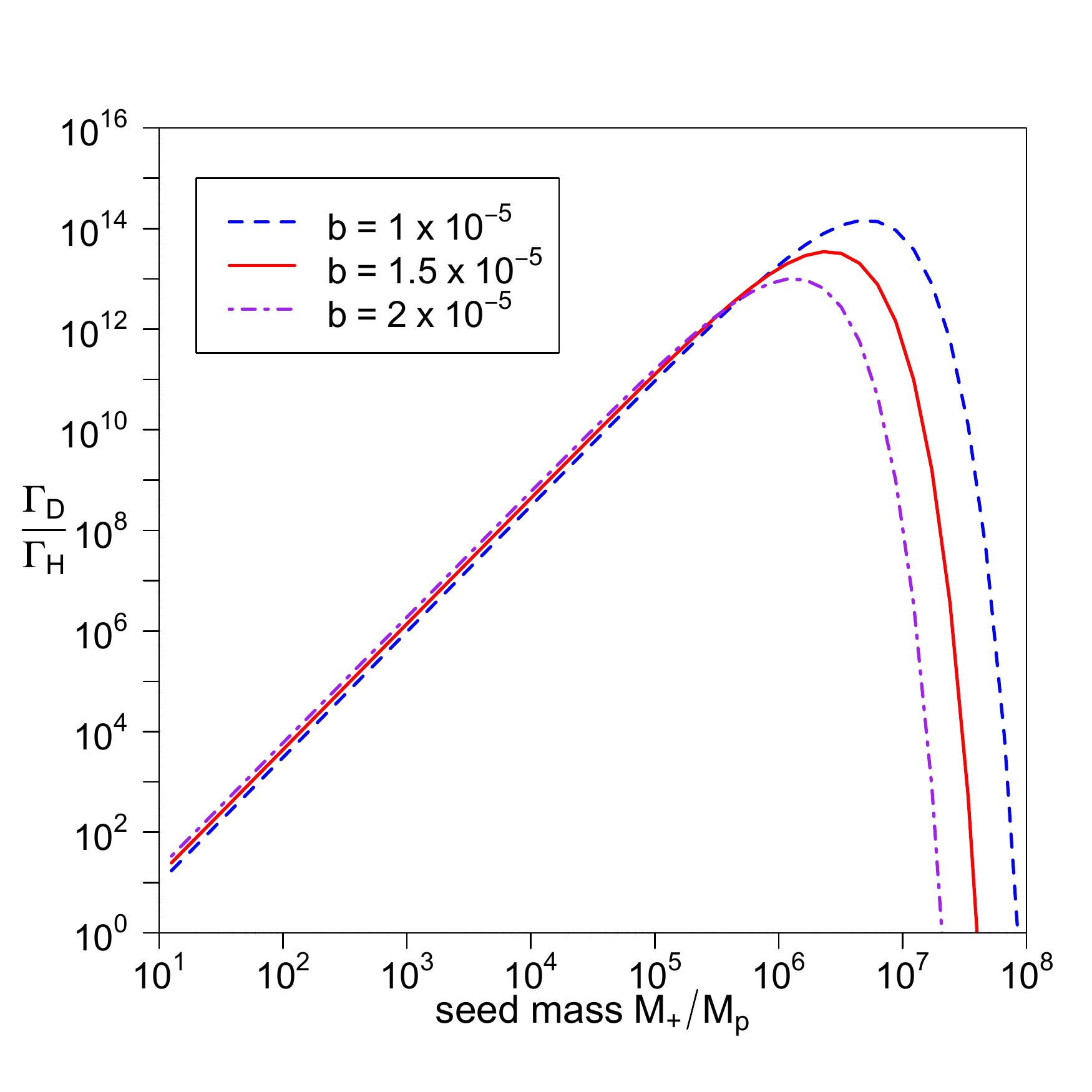}~
\includegraphics[scale=0.2]{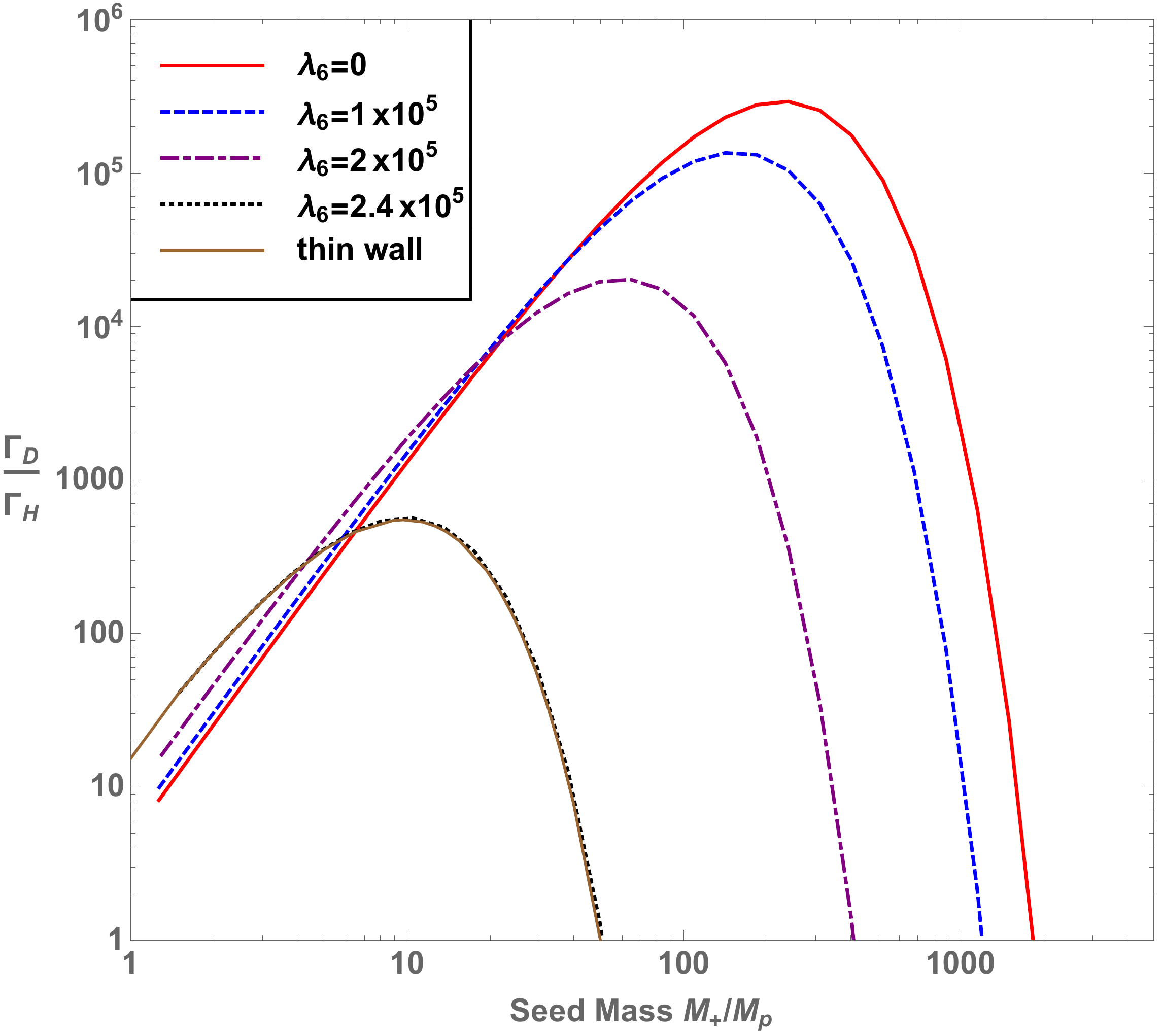}
\caption{The dependence of the branching ratio on the various model
parameters. On the left, the dependence of ratio on $\lambda_*$, in the middle, 
on $b$ and on the right the dependence on a quantum gravity motived 
$\lambda_6$ term showing the transition from thick to the analytic thin wall. 
\label{fig:branch}
}
\end{figure}

For the pure SM running, there is no thin wall bubble, and instantons must be found 
by numerical integration. In order to get a second minimum and make contact with 
the thin wall results of the previous section, one typically adds quantum gravity 
motivated terms such as $\frac{\lambda_6 \phi^2}{6 M_p^2}$
\cite{Donoghue:1994dn,Greenwood:2008qp}, with increasing
$\lambda_6$ taking us to the thin wall limit (see figure \ref{fig:branch}).
Motivated by the thin wall results, we search for a static bounce solution to the 
Euclidean Einstein-scalar equations on a black hole background. These are found
by using a spherically symmetric Ansatz for the metric 
\begin{equation}
ds^2=f(r)e^{2\delta(r)}d\tau^2+{dr^2\over f(r)}
+r^2(d\theta^2+\sin^2\theta d\varphi^2),
\quad
{\rm where}
\quad
f=1-{2G\mu(r)\over r}.
\end{equation}
and integrating the scalar and metric equations numerically (see \cite{BGM3}
for full details). Given that the instanton is static, the action is given by
the difference in area between the seed and remnant black holes. We then
input this action in the branching ratio (\ref{branchratio}) to determine the
relative risk for vacuum decay as shown in figure \ref{fig:branch}.

What figure \ref{fig:branch} shows is that for small black holes that are nonetheless
sufficiently larger than the Planck mass so that the semiclassical approximation is
valid, vacuum decay is enormously dominant over evaporation. 

\section{Conclusion}

We have demonstrated by analytical modelling backed up by numerical
results, that the lifetime of our universe is crucially dependent on the existence of
tiny inhomogeneities: the presence of even a single tiny black hole in our 
Hubble volume could trigger nucleation to a universe 
with a very different ``standard model''. Such small black
holes could arise as relics from the primordial phase of the universe
\cite{Carr:1974nx}.
Although black holes produced in the early universe start out with 
masses many orders of magnitude higher than $M_p$, they gradually 
evaporate until they come into the range shown in figure \ref{fig:branch}. 
At this point, the tunneling half life becomes smaller
than the (instantaneous) Hawking lifetime of the black hole: 
$\sim 10^{-28}$s for a $10^5M_p$ mass black hole! It is 
clear that once a primordial black hole nears the end of its life cycle, 
it {\it will} seed vacuum decay.

Another scenario where small black holes arise  is as a by-product of
collisions at the LHC \cite{Dimopoulos:2001hw} in some large extra 
dimension models. Such black holes have some higher-dimensional
features, notably a different entropic relation, and a preliminary analysis
\cite{BGM3} suggests they may be less problematic, though we are
exploring this in more detail.

Our conclusions of course depend on the existence of 
small black holes, and on the running of the Higgs coupling beyond 
standard model scales -- thus, our continued existence suggests either
that there are no primordial black holes, or that physics beyond the SM
stabilises the Higgs potential. Either way,
these results show that the issue of metastability of 
our universe may not be as simple as was initially thought.

\section*{Acknowledgments}

We would like to thank Philipp Burda and Ben Withers for collaboration on 
this project.


\begin{thebibliography}{99}

\bibitem{GMW}
R.~Gregory, I.~G.~Moss and B.~Withers,
JHEP {\bf 1403} (2014) 081,

\bibitem{BGM1} 
P.~Burda, R.~Gregory and I.~Moss,
Phys.\ Rev.\ Lett.\  {\bf 115}, 071303 (2015)

\bibitem{BGM2} 
P.~Burda, R.~Gregory and I.~Moss,
JHEP {\bf 1508}, 114 (2015)

\bibitem{BGM3}
P.~Burda, R.~Gregory and I.~Moss,
JHEP {\bf 1606}, 025 (2016)

\bibitem{ATLAS:2012ae} 
G.~Aad {\it et al.}  [ATLAS Collaboration],
Phys.Lett. {\bf B710}, 49 (2012)
 
\bibitem{Chatrchyan:2012tx} 
S.~Chatrchyan {\it et al.}  [CMS Collaboration],
Phys.Lett. {\bf B710}, 26 (2012)

\bibitem{Degrassi:2012ry} 
G.~Degrassi, S.~Di Vita, J.~Elias-Miro, J.~R.~Espinosa, 
G.~F.~Giudice, G.~Isidori and A.~Strumia,
JHEP {\bf 1208} (2012) 098,

\bibitem{Kobzarev:1974cp} 
I.~Y.~Kobzarev, L.~B.~Okun and M.~B.~Voloshin,
Sov.J.Nucl.Phys. {\bf 20} (1975) 644,
[Yad.Fiz. {\bf 20} (1974) 1229].

\bibitem{Coleman:1977py}
S.~Coleman, 
{\em  Phys.Rev.} {\bf D15} (1977) 2929--36.

\bibitem{Callan:1977pt}
C. G. Callan and S.~Coleman, 
{\em  Phys.Rev.} {\bf D16} (1977) 1762--68.

\bibitem{Coleman:1980aw}
S.~Coleman and F.~De~Luccia, 
{\em Phys. Rev. D} {\bf 21} (Jun, 1980) 3305--3315.

\bibitem{Israel}
W.~Israel,
Nuovo Cimento Soc. Ital. Phys. {\bf B44} (1966) 4349.

\bibitem{BCG}
P.~Bowcock, C.~Charmousis and R.~Gregory,
{\em Class.\ Quant.\ Grav.\ } {\bf 17}, 4745 (2000)

\bibitem{Donoghue:1994dn} 
J.~F.~Donoghue,
Phys.\ Rev.\ D {\bf 50}, 3874 (1994)

\bibitem{Greenwood:2008qp} 
E.~Greenwood, E.~Halstead, R.~Poltis and D.~Stojkovic,
Phys.Rev. {\bf D79} (2009) 103003,

\bibitem{Page:1976df}
D. N. Page, 
{\em  Phys.Rev.} {\bf D13} (1976) 198-206.

\bibitem{Carr:1974nx} 
B.~J.~Carr and S.~W.~Hawking,
Mon.\ Not.\ Roy.\ Astron.\ Soc.\  {\bf 168}, 399 (1974).

\bibitem{Dimopoulos:2001hw} 
S.~Dimopoulos and G.~L.~Landsberg,
Phys.\ Rev.\ Lett.\  {\bf 87}, 161602 (2001)

\end{thebibliography}
\end{document}